\documentclass[aps,prl,twocolumn,letterpaper,amsmath,amssymb,longbibliography]{revtex4-2}

\usepackage{graphicx}
\usepackage{dcolumn}
\usepackage{bm}
\usepackage{float}
\usepackage[T1]{fontenc}
\usepackage{txfonts}
\usepackage{fontawesome} 
\usepackage{xcolor}
\usepackage{hyperref}
\DeclareGraphicsExtensions{.eps,.pdf,.png,.jpg}

\definecolor{ExpBlue}{RGB}{31, 127, 180}

\definecolor{MeasRed}{RGB}{214, 39, 40}

\definecolor{SatOrange}{RGB}{255, 127, 14}

\definecolor{TpaPurple}{RGB}{148, 103, 189}

\newcommand{\ExpMarker}[1][]{open circle#1}
\newcommand{\AdjMarker}[1][]{open square#1}
\newcommand{\MeasMarker}[1][]{open diamond#1}
\newcommand{\SatMarker}[1][]{open triangle#1}

\newcommand{\kcsixty}{\ensuremath{\text{K}_{3}\text{C}_{60}}}
\newcommand{\funit}{\ensuremath{\text{mJ}/\text{cm}^2}}
\newcommand{\nsurf}{\ensuremath{n_\text{s}}}
\newcommand{\sigmasurf}{\ensuremath{\sigma_\text{s}}}
\newcommand{\sigmasurfre}{\ensuremath{\sigma_{\text{s}1}}}
\newcommand{\sigmasurfim}{\ensuremath{\sigma_{\text{s}2}}}
\newcommand{\nbar}{\ensuremath{\bar{n}}}
\newcommand{\rbar}{\ensuremath{\bar{r}}}
\newcommand{\Rbar}{\ensuremath{\bar{R}}}
\newcommand{\sigmabar}{\ensuremath{\bar{\sigma}}}
\newcommand{\Radj}{\ensuremath{R_{\text{adj}}}}
\newcommand{\epsinf}{\ensuremath{\epsilon_{\infty}}}
\newcommand{\sigmasat}{\ensuremath{\sigma_\text{sat}}}
\newcommand{\sigmatpa}{\ensuremath{\sigma_\text{TPA}}}
\newcommand{\fsat}{\ensuremath{F_{\text{sat}}}}
\newcommand{\ftpa}{\ensuremath{F_{\text{TPA}}}}
\newcommand{\pexp}{\ensuremath{\mathcal{P}_\text{exp}}}
\newcommand{\psat}{\ensuremath{\mathcal{P}_\text{sat}}}
\newcommand{\ptpa}{\ensuremath{\mathcal{P}_\text{TPA}}}
\newcommand{\deff}{\ensuremath{d_\text{eff}}}
\newcommand{\latt}{\ensuremath{\Lambda}}
\newcommand{\Gsq}{\ensuremath{G_\Box}}
\newcommand{\Tc}{\ensuremath{T_\text{c}}}
\newcommand{\ybco}{YBa\textsubscript{2}Cu\textsubscript{3}O\textsubscript{$y$}}

\begin{document}

\title{Optical Saturation Produces Spurious Evidence for\\ Photoinduced Superconductivity in \kcsixty}

\author{J. Steven Dodge}
\author{Leya Lopez}
\author{Derek G. Sahota}
\affiliation{Department of Physics, Simon Fraser University, Burnaby, British Columbia V5A~1S6, Canada}

\date{\today}

\begin{abstract}
We discuss a systematic error in time-resolved optical conductivity measurements that becomes important at high pump intensities. We show that common optical nonlinearities can distort the photoconductivity depth profile, and by extension distort the photoconductivity spectrum. We show evidence that this distortion is present in existing measurements on \kcsixty, and describe how it may create the appearance of photoinduced superconductivity where none exists. Similar errors may emerge in other pump-probe spectroscopy measurements, and we discuss how to correct for them.
\end{abstract}

\maketitle

A series of experiments over the last decade suggests that intense laser pulses may induce superconductivity in several materials~\cite{cavalleri2018, dong2022}. Time-resolved terahertz spectroscopy has supplied the main evidence for this effect, since it has the electrodynamic sensitivity and the subpicosecond time resolution necessary to observe its evolution~\cite{jepsen2011, ulbricht2011}. These measurements are commonly reported in terms of the complex photoexcited surface conductivity \mbox{$\sigmasurf = \sigmasurfre + i\sigmasurfim$}, which is derived from experiment as a function of frequency $\omega$ and pump-probe time delay $\Delta t$ using a standard analysis procedure~\cite{jepsen2011, ulbricht2011, orenstein2015}. Here, we show that this procedure distorts $\sigmasurf(\omega)$ when the photoinduced response has a nonlinear dependence on pump fluence---precisely the regime in which photoinduced superconductivity has been reported. As an example, we describe how the evidence for photoinduced superconductivity in \kcsixty~\cite{mitrano2016, cantaluppi2018, budden2021, buzzi2021} is susceptible to these distortions, and we present an alternative explanation for the results that does not involve superconductivity.

\begin{figure}[tbp]
\begin{center}
\includegraphics[width=\columnwidth]{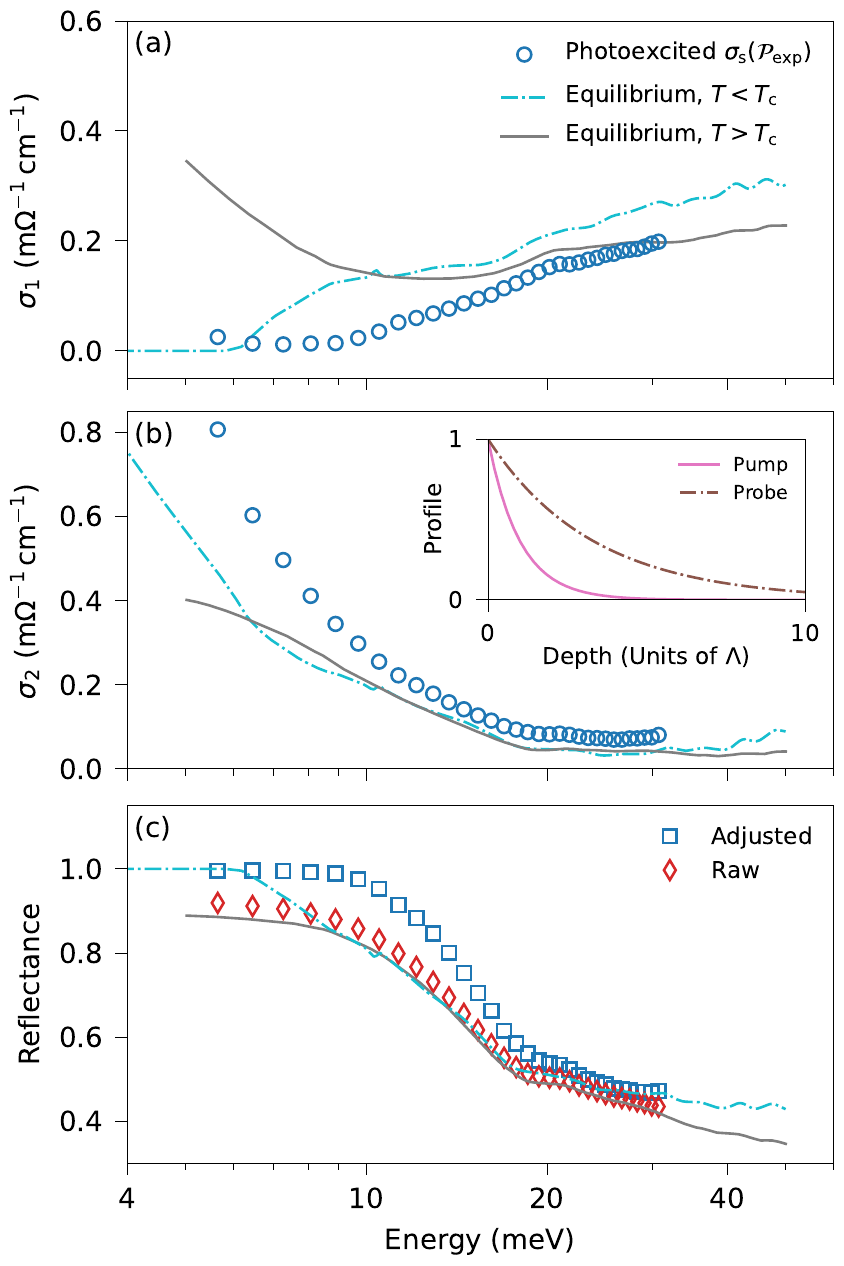}
\caption{Conductivity (a,b) and reflectance (c) of \kcsixty, in equilibrium and after photoexcitation with fluence \mbox{$F=3.0~\funit$} and pump photon energy \mbox{$\hbar\omega \approx 170$~meV}, adapted from~\mbox{\textcite{budden2021}}. Equilibrium results are shown above and below \mbox{$\Tc = 20~\text{K}$}, at 100~K and 10~K, respectively~\cite{cantaluppi2018}. Photoexcited results are shown at 100~K for \mbox{$\Delta t = 1$~ps}. The photoexcited surface conductivity \sigmasurf~(\ExpMarker[s]) is inferred by assuming the profile \pexp\ with \mbox{$\epsinf = 5$} and  \mbox{$\latt = 220$~nm}~\cite{mitrano2016, budden2021}. The distinction between the adjusted reflectance~(\AdjMarker[s]) and the raw reflectance~(\MeasMarker[s]) is described in the text. The inset shows $\mathcal{E}(z)$ for the pump and the probe in the linear optical regime at \mbox{$\hbar\omega = 6.46$~meV}, each normalized to their surface value.}
\label{fig:budden-summary}
\end{center}
\end{figure}

At equilibrium, the electrodynamic response of \kcsixty\ exhibits the characteristic features of a superconductor in the dirty limit, as shown in Fig.~\ref{fig:budden-summary}. The equilibrium complex conductivity \mbox{$\sigmabar = \sigmabar_1 + i\sigmabar_2$} above the critical temperature \Tc\ can be described by a semiclassical Drude-Lorentz model~\cite{buzzi2021}, where the Lorentz oscillators account for the broad mid-infrared conductivity at \mbox{$\hbar\omega \gtrsim 10$~meV} and the Drude response dominates at lower frequencies. (We use an overbar to distinguish static, equilibrium quantities from their time-dependent, nonequilibrium counterparts.) Below \Tc\ a gap opens in $\sigmabar_1$ at \mbox{$\hbar\omega\lesssim 6$~meV}, as spectral weight condenses into the superconducting \mbox{$\delta$ function} at \mbox{$\omega=0$}. Over the same frequency range, the equilibrium reflectance is lossless, with \mbox{$\Rbar = 1$}, and the inertial response of the superfluid causes $\sigmabar_2$ to diverge as $1/\omega$.
%
%

The optical properties reported for the photoexcited state with \mbox{$T>\Tc$} at \mbox{$\Delta t = 1$~ps} are qualitatively similar to those of the equilibrium superconducting state. At low frequencies, photoexcitation suppresses $\sigmasurfre$, enhances $\sigmasurfim$, and, after an adjustment that we discuss below, causes the reported reflectance \Radj\ to approach unity. The evidence for photoinduced superconductivity in \kcsixty\ hinges on these similarities~\cite{mitrano2016, cantaluppi2018, budden2021, buzzi2021}.

But there is a crucial difference between the two sets of measurements. The equilibrium conductivity is spatially uniform, so for a given background relative permittivity \epsinf\ there is a unique mapping from the measured complex reflection amplitude $\rbar$ to the quantity of interest, \sigmabar. This is not the case for the photoinduced response, since the photoconductivity $\Delta\sigma$ is not uniform. To determine \sigmasurf\ uniquely from the photoexcited reflection amplitude $r$, we must also specify the conductivity profile $\mathcal{P}$ as a function of the depth $z$ from the surface. If $\mathcal{P}$ is not known independently, we must assume a model for it. Any error in this model will be passed on to \sigmasurf.

Following previous practice~\cite{kaiser2014, hunt2016, mitrano2016, cantaluppi2018}, \mbox{\textcite{budden2021}} use a profile that we denote by \pexp, which they express in terms of the refractive index as
%
%
\begin{equation}
n(\omega, z; \pexp) = \nbar(\omega) + \Delta \nsurf(\omega)\,e^{-\alpha z},\label{eq:index-profile}
\end{equation}
where $\nbar$ is the equilibrium refractive index, $\alpha$ is the pump attenuation coefficient, and $\Delta \nsurf$ is the photoinduced change in the refractive index at the surface. We include the label \pexp\ explicitly to emphasize its role in inferring $n(\omega, z; \pexp)$ from the measured $r(\omega)$. In terms of the conductivity, the profile is
%
%
\begin{equation}
\begin{split}
\sigma(\omega, z; \pexp) &= \sigmabar(\omega) + \Delta\sigma(\omega, z;\pexp),\\
&= -i\omega\epsilon_0\{[n(\omega, z; \pexp)]^2 - \epsinf\}.
\end{split}\label{eq:exp-profile}
\end{equation}
Now it is possible to determine \mbox{$\sigmasurf(\omega; \pexp) = \sigma(\omega, 0;\pexp)$} from $r(\omega)$ by solving the Maxwell equations with \pexp\ and matching the usual electromagnetic boundary conditions at the surface~\cite{born1999}.

The problem with this procedure is that \pexp\ implicitly relies on two assumptions that are both unreliable. First, it assumes that the pump absorption remains linear in the pump intensity, so that the energy density $\mathcal{E}$ absorbed by the pump decays as \mbox{$\mathcal{E}\propto e^{-\alpha z}$}. Second, it assumes that $n$ is linear in $\mathcal{E}$. Jointly, these assumptions imply that Eq.~\eqref{eq:exp-profile} is independent of the pump intensity. But none of these assumptions are sound at the high pump intensities used in the experiments. Indeed, the measured photoresponse consistently shows a \emph{nonlinear} dependence on the incident fluence $F$~\cite{fausti2011, kaiser2014, nicoletti2014, casandruc2015, khanna2016, mitrano2016, cremin2019, liu2020, zhang2020, buzzi2020, buzzi}, so analyzing them in terms of the profile \pexp\ is not self-consistent.

And as we demonstrate here, neglecting nonlinearity can introduce errors in $\sigmasurf(\omega;\pexp)$ that are profoundly misleading. The pump attenuation length \mbox{$\latt = 1/\alpha$} is less than a third of a typical probe attenuation length in \kcsixty\ (see inset to Fig.~\ref{fig:budden-summary}), so the pump excites only a fraction of the probe volume and the photoinduced change in $r$ is much weaker than it would be with uniform excitation. The change $\Delta r$ is then mainly sensitive to the sheet photoconductance, \mbox{$\Delta\Gsq = \Delta\sigmasurf \deff$}, where \mbox{$\deff = \int dz\,\Delta\sigma(z)/\Delta\sigmasurf$} is the effective perturbation thickness. For \pexp, we get \mbox{$\deff = \latt$}, independent of fluence. But this is no longer true if the photoconductivity is nonlinear, and failing to account for this will introduce error in \deff. Any error in \deff\ will introduce a compensating error in $\Delta\sigmasurf$, distorting \sigmasurf.

The difference between the raw and adjusted reflectance in Fig.~\ref{fig:budden-summary}(c) reveals the scope for such an error. \mbox{\textcite{budden2021}} do not report raw measurements of the photoexcited reflectance \mbox{$R = |r|^2$}, so we have deduced it from their reported \pexp\ and \sigmasurf. What \mbox{\textcite{budden2021}} do report is \Radj, which they compute for an interface between a diamond window (used in the measurements) and a fictitious medium with uniform $\sigma(\omega)$ that they set equal to $\sigmasurf(\omega; \pexp)$. While the raw reflectance $R$ exceeds $\Rbar$ by at most 3.4\%, \Radj\ exceeds it by as much as 15\%, a discrepancy of more than a factor of 4. Note that $\Radj(\omega)$ is derived from $\sigmasurf(\omega; \pexp)$, not the other way around, so any error in \sigmasurf\ will also appear in \Radj. If we overestimate \deff, we will underestimate both $|\Delta\sigmasurf|$ and $|\Radj - \Rbar|$, and if we underestimate \deff\ we will overestimate them.
%
%
\begin{figure}[tbp]
\begin{center}
\includegraphics[width=\columnwidth]{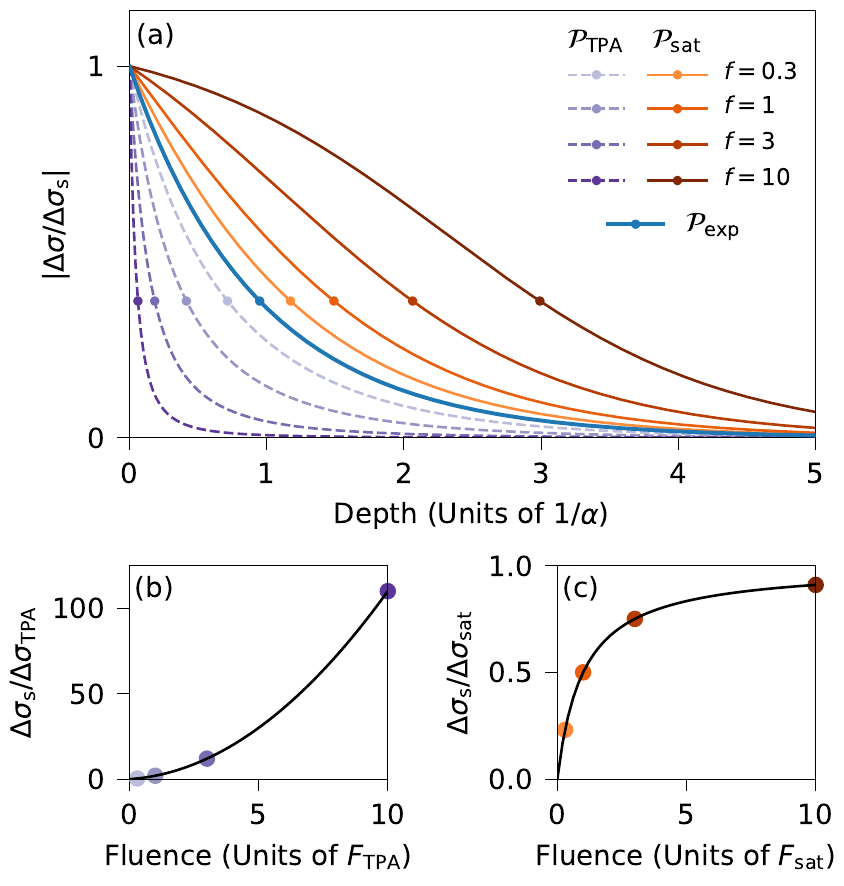}
\caption{Local photoconductivity $\Delta\sigma$ as a function of depth from the surface (a) and pump fluence (b),(c) for two models of nonlinearity. (a)~The profiles \psat\ (orange, solid lines) and \ptpa\ (purple, dashed lines) are shown for the same four values of the normalized pump fluence $f$, indicated by markers of the corresponding color in (b) and (c) for \ptpa\ and \psat, respectively. The profile \pexp\ (blue thick line) used by \mbox{\textcite{budden2021}} is shown for \mbox{$\hbar\omega = 6.46$~meV}. Markers in (a) indicate the $1/e$ depth for each curve.}
\label{fig:sat-profile}
\end{center}
\end{figure}

And as Fig.~\ref{fig:sat-profile} makes clear, nonlinearity can cause \deff\ to change by an order of magnitude or more as the fluence increases. We show profiles for two common nonlinearities, which we discuss in more detail in the Supplemental Material~\cite{supplement}. In one, which we label as \psat, we assume that \mbox{$\mathcal{E}\propto e^{-\alpha z}$} and that the local photoconductivity $\Delta\sigma$ saturates with $\mathcal{E}$. Defining the dimensionless fluence parameter \mbox{$f=F/\fsat$}, where \fsat\ is the characteristic scale for saturation, we express $\sigma$ as~\cite{petersen2017, sahota2019}
%
%
\begin{equation}
\sigma(\omega, z, f; \psat) = \bar{\sigma}(\omega) + \Delta\sigmasat(\omega)\frac{f e^{-\alpha z}}{1 + f e^{-\alpha z}}, \label{eq:sat-profile}
\end{equation}
which yields
%
%
\begin{equation}
\Delta\Gsq(\omega, f; \psat) = \Delta\sigmasat(\omega)\latt \ln(1+f)\label{eq:Gsq-sat}.
\end{equation}
Note that $\Delta\Gsq(\omega, f; \psat)$ continues to increase with $f$ even as \mbox{$\Delta\sigmasurf(\omega, f; \psat) = \Delta\sigmasat(\omega)f/(1+f)$} saturates. This is because $\Delta\sigma$ grows more slowly at the surface than it does in the interior as $f$ increases, which causes \deff\ to increase also. The logarithmic growth of $\Gsq(\omega, f; \psat)$ with $f$ does not depend on the detailed form of the saturation in Eq.~\eqref{eq:sat-profile}, since it follows from the assumption that \mbox{$\mathcal{E}\propto e^{-\alpha z}$}.

For the second profile, $\ptpa$, we assume that $\Delta\sigma$ remains proportional to $\mathcal{E}$ but that the absorption is nonlinear, with a two-photon absorption (TPA) coefficient $\beta$~\cite{supplement}. For simplicity, we further assume that the pump intensity has a rectangular temporal profile with duration $\tau_\text{p}$ and that the pump reflection coefficient $R_\text{p}$ remains constant. This allows us to express $\sigma$ analytically as
%
%
\begin{equation}
\sigma(\omega, z, f; \ptpa) = \bar{\sigma}(\omega) + \Delta\sigmatpa(\omega)\frac{f (1 + f) e^{-\alpha z}}{\left[1 + f(1-e^{-\alpha z})\right]^2},\label{eq:tpa-profile}
\end{equation}
where now \mbox{$f = F/\ftpa$} with \mbox{$\ftpa = (\alpha\tau_\text{p}/\beta)/(1 - R_\text{p})$}. As Fig.~\ref{fig:sat-profile}(b) shows, the surface photoconductivity \mbox{$\Delta\sigmasurf(\omega, f; \ptpa) = f(1+f)\Delta\sigmatpa(\omega)$} increases quadratically with $f$ when $f\gg 1$, where TPA dominates. At the same time, $\deff = \latt/(1+f)$ decreases with $f$, which compensates for the superlinear growth of $\Delta\sigmasurf$ and causes the sheet photoconductance,
%
%
\begin{equation}
\Delta\Gsq(\omega, f; \ptpa) = \Delta\sigmatpa(\omega)\latt f\label{eq:Gsq-tpa},
\end{equation}
to remain strictly proportional to $f$.

Now consider the systematic error that we introduce if we assume the wrong profile. If the true profile is \psat\ but we assume it is \pexp,  for example, then we would infer the surface conductivity to be \mbox{$\sigmasurf(\omega, f; \psat\mapsto\pexp)$}, where the notation \mbox{$\psat\mapsto\pexp$} indicates that we use the \psat\ profile to compute $r(\omega)$ with a source spectrum $\sigmasurf(\omega, f; \psat)$, then use the \pexp\ profile to infer an image spectrum $\sigmasurf(\omega; \psat\mapsto\pexp)$ from $r(\omega)$. The requirement that the source and image profiles yield the same $r(\omega)$ is roughly equivalent to holding \mbox{$\Delta\Gsq = \Delta\sigmasurf\deff$} constant for \kcsixty, so the image transformation effectively rescales the source $\Delta\sigmasurf$ by $\deff{(\mathcal{P}_\text{source})}/\deff{(\mathcal{P}_\text{image})}$. Since \mbox{$\deff = \latt$} for \pexp, we divide Eq.~\eqref{eq:Gsq-sat} by \latt\ to get
%
%
\begin{equation}
\Delta\sigmasurf(\omega, f; \psat\mapsto\pexp) \approx \Delta\sigmasat(\omega)\ln(1 + f),\label{eq:sat-exp-fluence}
\end{equation}
which overestimates $\Delta\sigmasurf(\omega, f; \psat)$ by \mbox{$(1+f)\ln(1+f)/f$}. Similarly, dividing Eq.~\eqref{eq:Gsq-tpa} by \latt\ gives
%
%
\begin{equation}
\Delta\sigmasurf(\omega, f; \ptpa\mapsto\pexp) \approx \Delta\sigmatpa(\omega)f,\label{eq:tpa-exp-fluence}
\end{equation}
which underestimates $\Delta\sigmasurf(\omega, f; \ptpa)$ by $1/(1+f)$.

\begin{figure}[tbp]
\begin{center}
\includegraphics[width=\columnwidth]{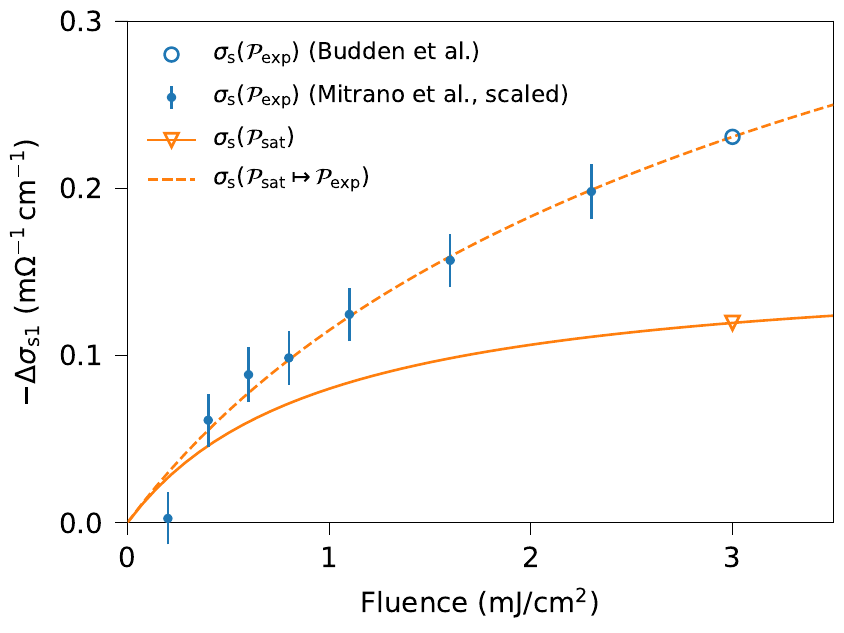}
\caption{Least-squares fit with \mbox{$\Delta\sigmasurfre(\omega, f; \psat\mapsto\pexp)$} (dashed line) to the fluence dependence of $\Delta\sigmasurfre(\omega; \pexp)$ reported by \mbox{\textcite{mitrano2016}} (points with error bars). The fit is constrained to pass through the anchor point $\Delta\sigmasurfre(\omega; \pexp)$~(\ExpMarker) at \mbox{$F=3~\funit$} reported by \mbox{\textcite{budden2021}} for \mbox{$\hbar\omega = 6.46$~meV}. We multiply the results of \mbox{\textcite{mitrano2016}} by an overall scale factor $A$ to account for systematic differences from the results of \mbox{\textcite{budden2021}}.  Best-fit parameter values are \mbox{$\fsat = (1.0 \pm 0.5)~\funit$} and \mbox{$A = 0.65 \pm 0.06$} (\mbox{$\chi^2 = 3.7$}, \mbox{$\text{d.o.f.} = 5$}). The solid line extrapolates the source function $\Delta\sigmasurfre(\omega, f; \psat)$ from its value of at \mbox{$F=3~\funit$}~(\SatMarker).}
\label{fig:fluence-dep}
\end{center}
\end{figure}

Figure~\ref{fig:fluence-dep} shows the fluence dependence reported by \mbox{\textcite{mitrano2016}} for $\Delta\sigmasurfre(\omega; \pexp)$ in \kcsixty, which we use to infer the profile. The measurements reveal a clear sublinear fluence dependence that is inconsistent with the relationship expected for \mbox{$\Delta\sigmasurfre(\omega, f; \ptpa\mapsto\pexp)$} given in Eq.~\eqref{eq:tpa-exp-fluence}~\cite{supplement}. And as we noted earlier, the deviation from linearity is also incompatible with the assumptions that yield the \pexp\ profile used in the original analysis. A fit with \mbox{$\Delta\sigmasurfre(\omega, f; \psat\mapsto\pexp)$}, however, is nearly indistinguishable from the experimental results---which means that the source function $\Delta\sigmasurfre(\omega, f; \psat)$, shown as a solid line in Fig.~\ref{fig:fluence-dep}, is the best estimate for the true surface photoconductivity. Note that this deviates significantly from the originally reported results at all fluences, and is nearly a factor of 2 smaller than the result reported by \mbox{\textcite{budden2021}} at \mbox{$F = 3.0~\funit$}.
%
%

\begin{figure}[tbp]
\begin{center}
\includegraphics[width=\columnwidth]{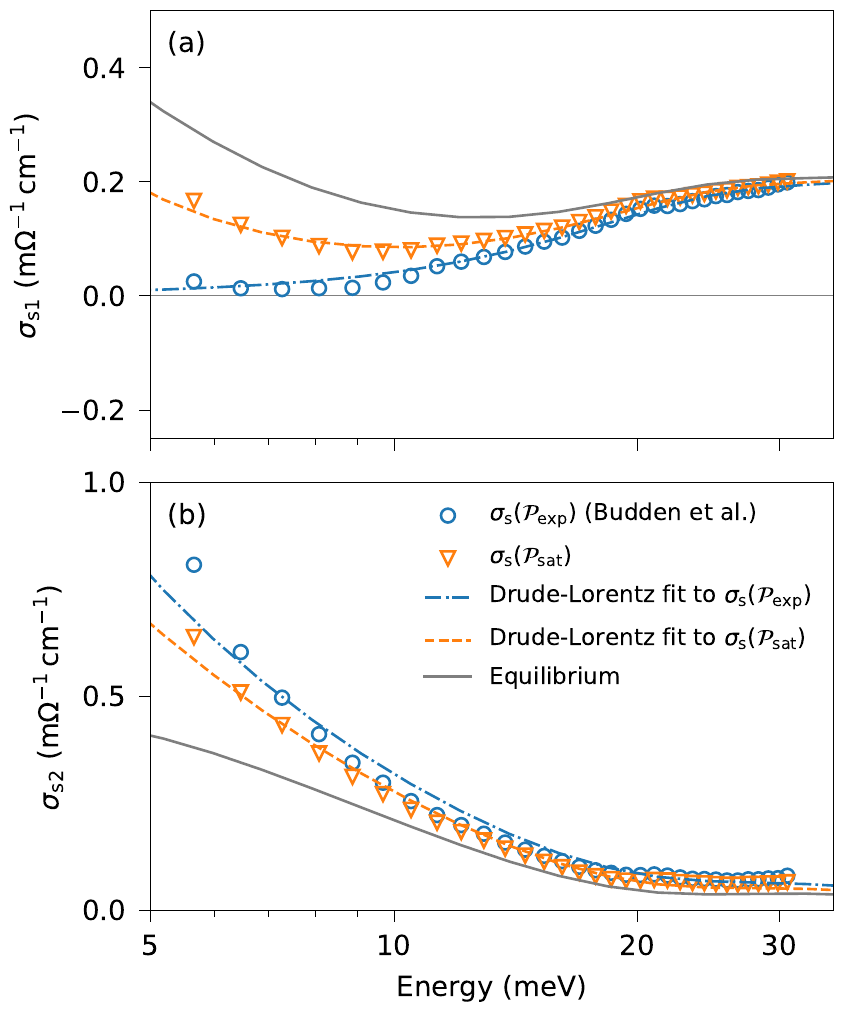}
\caption{Real (a) and imaginary (b) parts of \sigmasurf\ for \mbox{$F = 3.0~\funit$} with different profile assumptions. The spectrum $\sigmasurf(\omega; \pexp)$~(\ExpMarker[s]) reported by \mbox{\textcite{budden2021}} and the alternative spectrum \mbox{$\sigmasurf(\omega, f; \psat)$~(\SatMarker[s])} yield the same $r(\omega)$. Lines show Drude-Lorentz fits to $\sigmasurf(\omega; \pexp)$ (dot-dashed), $\sigmasurf(\omega, f; \psat)$ (dashed), and $\sigmabar(\omega)$ (solid)~\cite{supplement}.}
\label{fig:budden-alt}
\end{center}
\end{figure}

We fix \fsat\ at the value obtained from this fit and extend our analysis as a function of frequency in Fig.~\ref{fig:budden-alt}. We derive the alternative spectrum \mbox{$\sigmasurf(\omega; \psat)$} so that its image in \pexp\ is equal to $\sigmasurf(\omega; \pexp)$ reported by \mbox{\textcite{budden2021}}. Both spectra show decreases in $\sigmasurfre(\omega)$ and increases in $\sigmasurfim(\omega)$, but by different amounts. Since $\Delta\sigmasurf$ is inversely related to \deff, \mbox{$\Delta\sigmasurf(\omega; \pexp)$} has a smaller magnitude than \mbox{$\Delta\sigmasurf(\omega, f; \psat)$}.

These quantitative differences suggest qualitatively different physical interpretations. The spectrum with \pexp\ looks like that of a superconductor~\cite{mitrano2016, cantaluppi2018, budden2021}: $\sigmasurfre(\omega; \pexp)$ falls to near zero below \mbox{$\hbar\omega \approx 10$~meV}, $\sigmasurfim(\omega; \pexp)$ is enhanced at low frequencies, and a Drude-Lorentz fit yields a carrier relaxation rate \mbox{$\gamma = 0$}~\cite{supplement}. But the spectrum with \psat\ looks like a normal metal with a photoenhanced mobility: \mbox{$\sigmasurfre(\omega, f; \psat)$} lies well above zero at all $\omega$ and clearly increases with decreasing $\omega$ below \mbox{$\hbar\omega \approx 9$~meV}, while $\sigmasurfim(\omega, f; \pexp)$ shows more moderate enhancement at low frequencies. A Drude-Lorentz fit to this spectrum yields \mbox{$\hbar\gamma = 1.2$~meV}~\cite{supplement}, which is about a third of the equilibrium value~\cite{buzzi2021} and 4 times larger than the previously reported upper bound~\cite{cantaluppi2018}.
%
%

\begin{figure}[tbp]
\begin{center}
\includegraphics[width=\columnwidth]{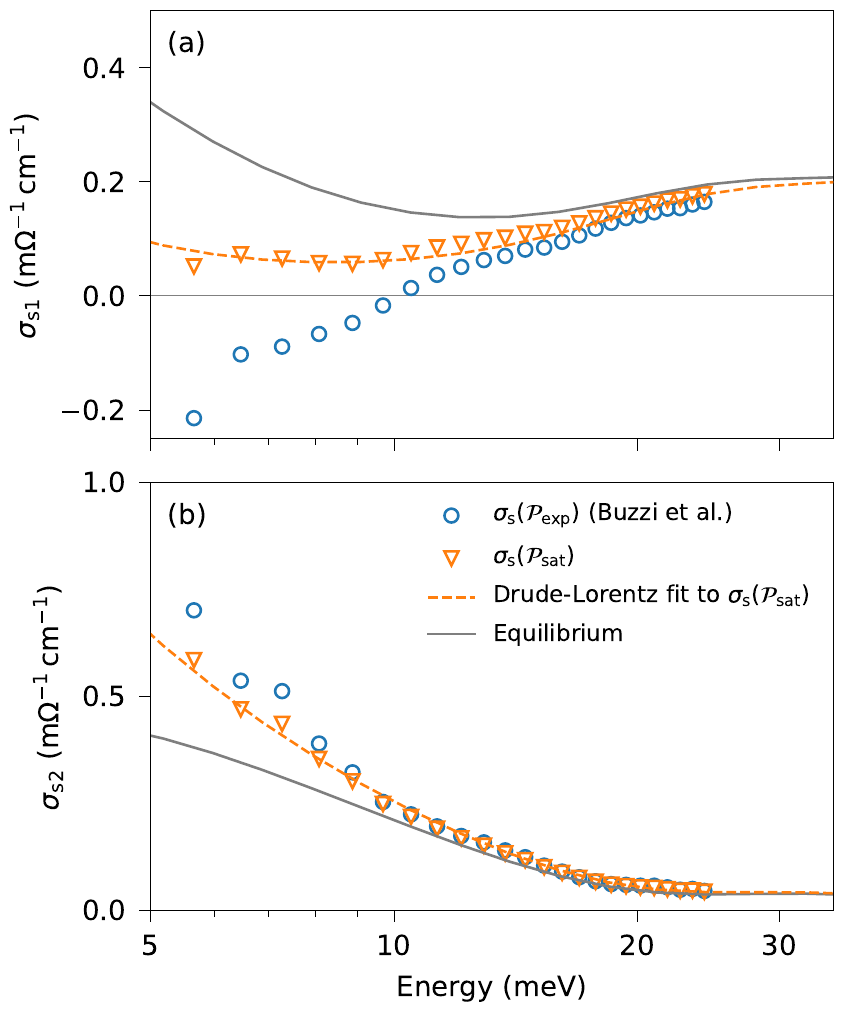}
\caption{Profile dependence of the real (a) and imaginary (b) parts of \sigmasurf\ reported by \mbox{\textcite{buzzi2021}} for \mbox{$F = 4.5~\funit$}. Markers and lines correspond to those in Fig.~\ref{fig:budden-alt}.}
\label{fig:buzzi-alt}
\end{center}
\end{figure}

We turn to measurements of \kcsixty\ at higher pump fluence for further guidance. Figure~\ref{fig:buzzi-alt} shows $\sigmasurf(\omega; \pexp)$ at \mbox{$F=4.5~\funit$} reported by \mbox{\textcite{buzzi2021}}, along with the alternative spectrum \mbox{$\sigmasurf(\omega, f; \psat)$}, defined in the same way as in Fig.~\ref{fig:budden-alt}. The higher fluence produces larger changes in \sigmasurf\ with both profiles, driving \sigmasurfre\ negative for $\sigmasurfre(\omega; \pexp)$. \mbox{\textcite{buzzi2021}} interpreted this negative real conductivity as evidence for Higgs-mediated optical parametric amplification, generated by a rapid quench from a superconducting state. For this to work, the pump would need to both produce a transient superconducting state and quench it within 100~fs, since the experiments are conducted above the equilibrium \Tc. But when we assume \psat\ instead of \pexp, a simpler interpretation emerges. A Drude-Lorentz fit to \mbox{$\sigmasurf(\omega, f; \psat)$} yields \mbox{$\hbar\gamma = 0.6$~meV}, about half the value obtained for \mbox{$F=3.0~\funit$} and 1/6 the equilibrium value~\cite{supplement}. Neither photoinduced superconductivity nor Higgs-mediated amplification are necessary to explain the results. At all fluences, the measurements are consistent with a relatively moderate photoinduced enhancement of the carrier mobility.
%
%

While we have focused here on \kcsixty\ at the moment of peak response, our observations raise important interpretational questions about the entire body of experimental literature on photoinduced superconductivity. In \ybco, for example, a signal associated with photoinduced superconductivity appears to show a linear dependence on the peak electric field, but the dependence is also consistent with what we have described for a saturable medium~\cite{kaiser2014, liu2020, supplement}. Furthermore, this signal is enhanced when the pump is tuned to specific phonon resonances, but if the photoconductivity saturates more easily at these resonances, the signal enhancement could be caused by changes in \deff\  instead of $\Delta\sigmasurf$~\cite{kaiser2014, liu2020}. In fact, the first report of photoinduced superconductivity described a similar mechanism~\cite{fausti2011}. This work also  noted that both \deff\ and the signal strength should grow logarithmically with fluence as a result of the saturation, following reasoning similar to ours~\cite{fausti2011}. Subsequent work failed to incorporate these insights, however, and needs reassessment.

There are several ways to overcome the problems that we have identified. Measurements on thin films with thickness $t\lesssim\latt$ would be ideal, as they would eliminate the uncertainty in \deff. In principle, ellipsometric measurements could determine \sigmasurf\ and \deff\ simultaneously, although in practice this would be technically challenging. \mbox{\textcite{katsumi}} has used nonlinear THz measurements to test for the existence of photoinduced superconductivity in \ybco, and found none. Another approach is to examine the joint dependence of $r$ on frequency, fluence, and time to specify a parametrized model for \sigmasurf, as we have described here. All of these approaches would help us to decide if the photoinduced superconductivity observed in \kcsixty\ and other compounds is real---or if it is an artifact of nonlinear distortion.

\begin{acknowledgments}
We thank M.\ Hayden, J.\ Orenstein, and D.\ Broun for critical feedback. J.\,S.\,D.\ acknowledges support from NSERC and CIFAR, and D.\,G.\,S.\ from an NSERC Alexander Graham Bell Canada Graduate Scholarship.
\end{acknowledgments}
\bibliography{pisc}

\end{document}


\title{Optical Saturation Produces Spurious Evidence for Photoinduced Superconductivity in \kcsixty\\
\normalfont{Supplemental Material}}

\author{J. Steven Dodge}
\author{Leya Lopez}
\author{Derek G. Sahota}
\affiliation{Department of Physics, Simon Fraser University, V5A~1S6, Canada}

\date{\today}

\maketitle

\section{Optical nonlinearities}\label{sec:nlo}
Saturation and two-photon absorption (TPA) are common optical nonlinearities that are observed in atoms, molecules and solids, and our analysis assumes standard phenomenological forms for each of these effects~\cite{boyd2008}. While neither effect has been studied in detail in \kcsixty, we outline here some of the basic physical mechanisms that could produce them, and discuss related effects that could distort the photoconductivity spectrum in similar ways.

\subsection{Saturation due to physical constraints}\label{sec:sat}
We expect the optical response of a system to saturate whenever it approaches a physical constraint, such as $0\leq R \leq 1$ or $\sigma_1\geq 0$ for a system in quasiequilibrium. In the Drude-Lorentz model, the momentum relaxation rate must also satisfy $\gamma\geq 0$, which we consider in more detail in Sec.~\ref{sec:gamma-sat}. While the response of a nonequilibrium system with optical gain may violate these bounds, it will still be constrained by energy conservation and by the energy transfer process that produces the gain~\cite{loudon2000}.

We can motivate our specific model for saturation by postulating that photoexcitation causes the system to fluctuate on short length and time scales into a distinct  phase of matter, and that $\sigma(\omega, z, f; \psat)$ represents a local average over these fluctuations. In this case, saturation results from the physical constraint that the time-averaged volume fraction of the photoinduced phase can not exceed unity. Equation~(3) of the main text is a phenomenological assumption for this locally averaged $\sigma$, and is based on the simplest Pad{\'{e}} approximant that is linear in $f$ when $f\ll 1$ and saturates with $f$ as $f\rightarrow\infty$.

We may associate \fsat\ with a characteristic energy density for saturation, $\mathcal{E}_\text{sat} = (1- R_\text{p})\alpha\fsat$, where $R_\text{p}\approx 0.8$ is the pump reflection coefficient of \kcsixty\ at the pump wavelength~\cite{degiorgi1994a}. For our estimate of \mbox{$\fsat \approx 1.0~\funit$}, we get \mbox{$\mathcal{E}_\text{sat} \approx 50~\text{meV}/\text{nm}^3$}.

\subsection{Application to metastable states}\label{sec:metastable}
Similar considerations apply to metastable states at longer times~\cite{zhang2018, cremin2019, budden2021, sun2020}. Metastability implies that there is a barrier between two minima in the free energy, which should produce a nonlinear response to photoexcitation. The barrier will set an energy density threshold $\mathcal{E}_\text{c}$ to reach the metastable state, which will set a characteristic fluence scale for nonlinearity. Assuming \mbox{$\mathcal{E} = \mathcal{E}_0e^{-\alpha z}$}, the volume that can make the transition into the metastable state will be proportional to $\ln(\mathcal{E}_0/\mathcal{E}_\text{c})$ and grow logarithmically with increasing fluence, similar to the behavior that we have described for \psat. Ignoring the effect of saturation on the depth profile will amplify $\Delta\sigmasurf$ and can make a state with enhanced mobility look like a superconductor over a finite frequency range~\cite{budden2021}.

\subsection{Saturation due to heating}\label{sec:heat}
Heating can also distort the photoconductivity spectrum at longer time delays~\cite{niwa2019,perez-salinas2022}. Once a photoexcited system has thermalized, the local temperature $T$ will typically have a nonlinear dependence on fluence. Assuming a Debye model, for example, the specific heat is proportional to $T^3$, so the local temperature will increase by approximately $\Delta T \propto \mathcal{E}^{1/4}$~\cite{petersen2017, niwa2019}. Taking $\Delta\sigma\propto\Delta T$, $\sigma$ will have a strongly sublinear dependence on $F$, which produces a thermal profile \pth\ that is qualitatively similar to \psat.

In an experiment that examined how photoexcitation destroys superconductivity in \lsco, \mbox{\textcite{niwa2019}} showed that \pth\ provides a better description of their measurements than \pexp, and found that the image spectrum \mbox{$\sigmasurf(\omega, T;\pth\mapsto\pexp)$} displays artifacts that are reminiscent of the results reported for \kcsixty\ by \mbox{\textcite{buzzi2021}}, including \mbox{$\sigma_1 < 0$} and \mbox{$\Radj > 1$}. \mbox{\textcite{buzzi2021}} considered the role of heating in their measurements and found that it could not explain their observations. But as we have discussed, any nonlinearity will distort the photoconductivity obtained with \pexp, and heating is not the only way to produce nonlinearity. As we have shown, their observations are consistent with a saturable photoenhancement of the mobility.

\subsection{Two-photon absorption}\label{sec:tpa}
Our model of TPA assumes that the pump intensity $I$ satisfies
\begin{equation}\label{eq:tpa-ode}
\frac{dI}{dz} = -\alpha I - \beta I^2,
\end{equation}
where $\alpha$ is the linear (i.e., one-photon) absorption coefficient, $\beta$ is the TPA coefficient, and we ignore higher-order absorption processes. This has the solution
\begin{equation}\label{eq:tpa-sol}
I(z, t) = \frac{I_\text{s}(t) e^{-\alpha z}}{1 + (\beta I_\text{s}/\alpha)(1 - e^{-\alpha z})},
\end{equation}
where $I_\text{s}(t) = I(z=0^+, t)$. Assuming for simplicity that the pump illumination is constant over the probe surface and has a rectangular temporal profile with pulsewidth $\tau_\text{p}$, the absorbed energy density is
\begin{equation}\label{eq:tpa-energy-density}
\mathcal{E}(z) = -\int_{-\tau_\text{p}/2}^{\tau_\text{p}/2} dt\,\frac{dI}{dz} = (1 - R_\text{p})\alpha \ftpa \frac{f(1+f)e^{-\alpha z}}{1 + f(1 - e^{-\alpha z})},
\end{equation}
where \mbox{$\ftpa = (\alpha\tau_\text{p}/\beta)/(1 - R_\text{p})$} and $f=F/\ftpa$.

Using the parameters for the \kcsixty\ measurements and setting \mbox{$\beta \approx 30~\text{cm}/\text{GW}$}, the value for GaAs at \mbox{$\lambda = 1064~\text{nm}$}~\cite{boyd2008}, we obtain an estimate of \mbox{$\ftpa\approx 750~\funit$}. This is nearly three orders of magnitude larger than the fluences used in the \kcsixty\ measurements, which supports our conclusion that TPA is not the dominant nonlinearity. However, as we discuss in the next section, we expect the same profile for excited-state absorption, and this could be significantly stronger than TPA~\cite{tutt1993}.

\subsection{Saturable absorption and excited-state absorption}\label{sec:sa-esa}
Our model of TPA assumes that the nonlinear response at the pump frequency is instantaneous, so that Eq.~\eqref{eq:tpa-ode} depends on the intensity $I$, not the absorbed energy density $\mathcal{E}$. Also, our model of saturation assumes that the optical response at the pump frequency remains linear in the pump field, so that the absorbed energy density retains the fixed form $\mathcal{E} \propto e^{-\alpha z}$. Saturable absorption (SA) and excited-state absorption (ESA) both violate these assumptions, but should produce distortions in the pump-probe response that are qualitatively similar to our models of saturation and TPA, respectively. Phenomenologically, we can describe both of these effects by assuming that the pump attenuation coefficient is a function of the absorbed energy density, $\alpha = \alpha(\mathcal{E})$, which yields the coupled differential equations
\begin{align}
\label{eq:esa-ode-}
\frac{\partial I}{\partial z} &= -\alpha(\mathcal{E})I, &
\frac{\partial \mathcal{E}}{\partial t} &= -\frac{\partial I}{\partial z} = \alpha(\mathcal{E})I.
\end{align}
We associate $d\alpha/d\mathcal{E} < 0$ with SA and $d\alpha/d\mathcal{E} > 0$ with ESA. We can combine these equations, expand $\alpha$ to first order in $\mathcal{E}$, and integrate over time to obtain a differential equation with the same structure as Eq.~\eqref{eq:tpa-ode}, but in terms of the fluence $F$~\cite{tutt1993},
\begin{equation}
\label{eq:esa-ode2}
\frac{dF}{dz} = -\alpha F - \gamma F^2,
\end{equation}
where the parameter $\gamma$ is the analogue of the TPA coefficient $\beta$. We emphasize that Eq.~\eqref{eq:esa-ode2} depends on the fluence, while Eq.~\eqref{eq:tpa-ode} depends on the instantaneous intensity, so while the equations have the same form, they depend differently on the pump pulse duration. (This distinction is also important in Sec.~\ref{sec:field-fluence-dep}.) For $\gamma > 0$, corresponding to ESA, Eq.~\eqref{eq:esa-ode2} yields the same excitation profile as for TPA, and \deff\ decreases with increasing fluence. With $\gamma < 0$, corresponding to SA, we expect \deff\ to increase with increasing fluence, as in our saturation model.

\section{Comparison of \mbox{$\Delta\sigmasurfre(\omega, f; \ptpa\mapsto\pexp)$} with experiment}\label{sec:tpa-exp-fluence}
\begin{figure}[tbhp]
\begin{center}
\includegraphics[width=0.45\columnwidth]{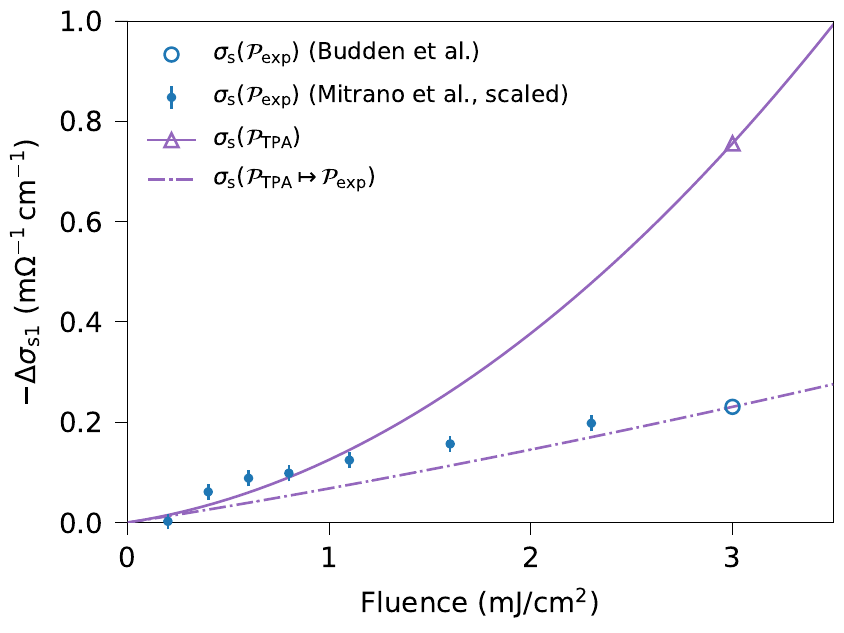}
\caption{Fluence dependence of \mbox{$\Delta\sigmasurfre(\omega, f; \ptpa\mapsto\pexp)$} (dot-dashed line) for \mbox{$\hbar\omega = 6.46$~meV} and \mbox{$\ftpa = 1.0~\funit$}, compared to $\Delta\sigmasurfre(\omega; \pexp)$ (points with errorbars) reported by \mbox{\textcite{mitrano2016}}. As in Fig.~3 of the main text, we constrain the value of \mbox{$\Delta\sigmasurfre(\omega, f; \ptpa\mapsto\pexp)$} to pass through the anchor point $\Delta\sigmasurfre(\omega; \pexp)$~(\ExpMarker) at \mbox{$F=3~\funit$} reported by \mbox{\textcite{budden2021}}, and we multiply the results of \mbox{\textcite{mitrano2016}} by the same overall scale factor $A = 0.65$ used in Fig.~3 of the main text. The solid line extrapolates the source function $\Delta\sigmasurfre(\omega, f; \ptpa)$ from its value at \mbox{$F=3~\funit$}~(\TpaMarker).}
\label{fig:fluence-tpa}
\end{center}
\end{figure}

Figure~\ref{fig:fluence-tpa} demonstrates the quality of the approximation \mbox{$\Delta\sigmasurf(\omega, f; \ptpa\mapsto\pexp)\approx\Delta\sigmatpa(\omega)f$} derived in the text. For the conditions of the measurements, we find that the exact fluence dependence is weakly superlinear for any finite value for \ftpa, which is inconsistent with the sublinear fluence dependence observed for $\Delta\sigmasurfre(\omega; \pexp)$.
%
%

\section{Drude-Lorentz fits}\label{sec:drude-lorentz}
For the spectral fits described in the main text, we follow \mbox{\textcite{buzzi2021}} and describe the equilibrium conductivity by the sum of a single Drude peak and a background composed of three Lorentzians,
\begin{equation}
\sigmasurf(\omega) = \sigma_\text{D}(\omega) + \sigma_\text{b}(\omega) = \frac{\epsilon_0\omega_\text{p}^2}{\gamma - i\omega} + \sum_{k=1}^3\frac{\epsilon_0\Omega_{\text{p}k}^2\omega}{i(\omega_{0k}^2 - \omega^2) + \Gamma_k\omega},
\end{equation}
with Drude parameters \mbox{$\hbar\omega_\text{p} = 160.61$~meV} and \mbox{$\gamma = 3.56$~meV} and the Lorentz parameters given in Table~\ref{tbl:lorentz-param}. Fits to the photoexcited conductivity hold the Lorentz parameters constant and allow the Drude parameters to vary.
%
%
\begin{table}[tbhp]
\caption{Lorentz parameters for \kcsixty\ in equilibrium~\cite{buzzi2021}.}
\begin{center}
\begin{ruledtabular}
\begin{tabular}{cccc}
$k$  & $\hbar\omega_{0k}$ (meV) & $\hbar\Omega_{\text{p}k}$ (meV) & $\hbar\Gamma_k$ (meV)\\
\colrule
1 & 26.1 & 184.92 & 34.0\\
2 & 70.4 & 368.84 & 86.6\\
3 & 102.6 & 133.57 & 35.0\\
\end{tabular}
\end{ruledtabular}
\end{center}
\label{tbl:lorentz-param}
\end{table}%

\section{Spectral amplitude in the saturation model}\label{sec:spectral-amp}
\begin{figure}[tbhp]
\begin{center}
\includegraphics[width=0.85\columnwidth]{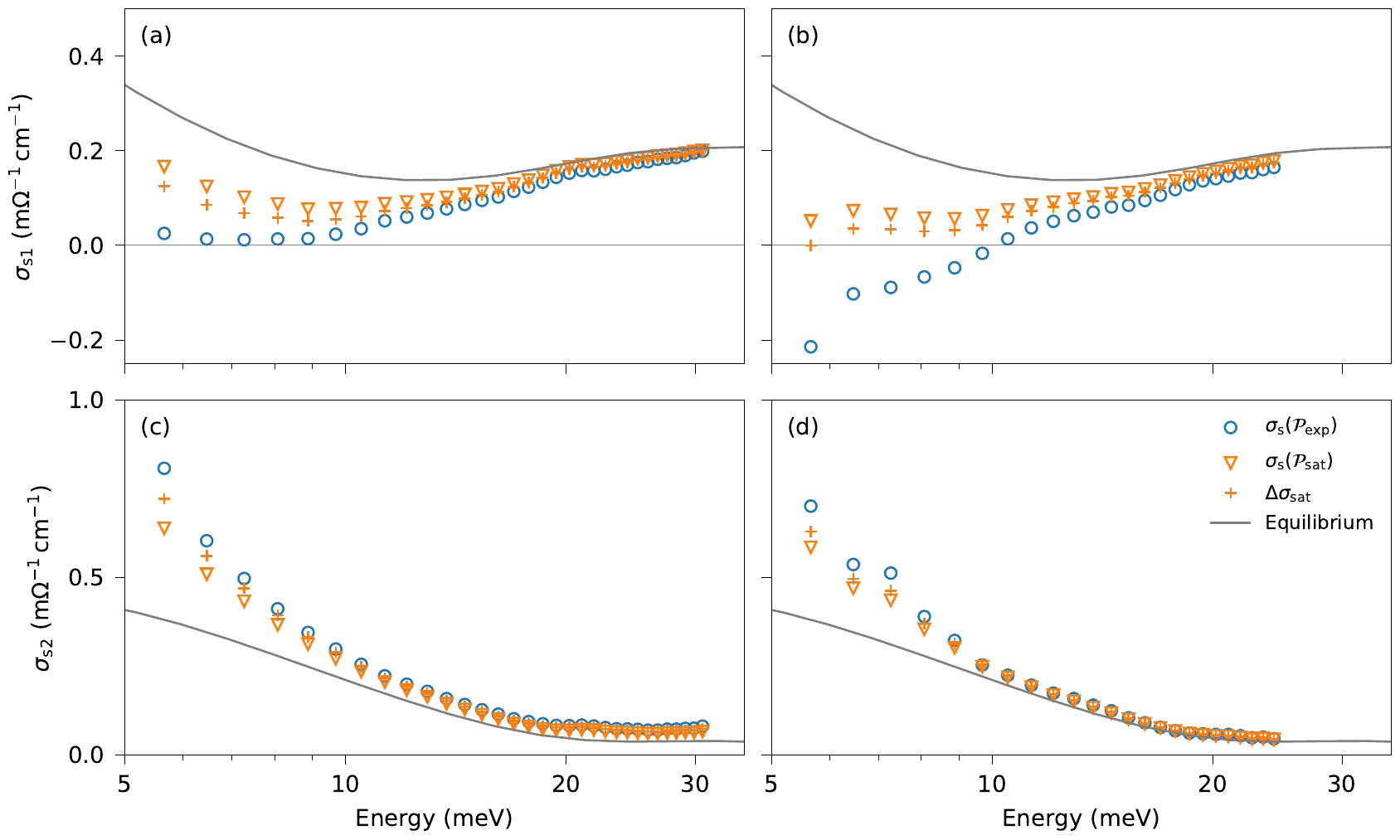}
\caption{Spectral amplitudes $\Delta\sigmasat(\omega)$~(\SigSatMarker) for the model profile \psat. For comparison, the real (a,b) and imaginary (c,d) parts of \sigmasurf\ for \mbox{$F = 3.0~\funit$}~(a,c)~\cite{budden2021} and for \mbox{$F = 4.5~\funit$}~(b,d)~\cite{buzzi2021} with different profile assumptions are reproduced from Fig.~4 and Fig.~5 of the primary text. Solid lines show $\sigmabar(\omega)$.}
\label{fig:spectral-amp}
\end{center}
\end{figure}

The spectral amplitude $\Delta\sigmasat(\omega)$ obtained for \mbox{$F=3.0~\funit$} varies from the one obtained at \mbox{$F=4.5~\funit$} by a maximum of 20\%, as shown in Fig.~\ref{fig:spectral-amp}. We tentatively attribute this variation to run-to-run measurement variability, but it could indicate a need for more flexibility in the model profiles. In principle, we could accommodate the variation by allowing $\Delta\sigmasat$ to depend on fluence as well as on frequency, but in practice this has little influence on \sigmasurf, since we have already accounted for the dominant fluence dependence in the model profile. Introducing fluence dependence here also risks overfitting to variation that could arise from changes in the experimental conditions instead of changes in the photoexcitation profile. We leave this function as frequency-independent in our presentation for simplicity.
%
%

\section{Saturable scattering rate profile}\label{sec:gamma-sat}
Here we consider a Drude-Lorentz model in which the carrier relaxation rate decreases with fluence and saturates at a minimum value $\gammasat\geq 0$. Letting \fgamma\ be the fluence scale for saturation and \mbox{$f = F/\fgamma$}, we have
\begin{gather}
\sigma(\omega, z, f; \mathcal{P}_\gamma) = \frac{\epsilon_0\omegap^2}{\gamma(z, f) - i\omega} + \sigma_\text{b}(\omega),  \label{eq:gamma-profile}\\
\gamma(z, f) =\gammabar + \Delta\gamma\frac{f e^{-\alpha z}}{1 + f e^{-\alpha z}}, \label{eq:rate-saturation-gamma}
\end{gather}
where \gammabar\ is the equilibrium scattering rate, \mbox{$\Delta\gamma = \gammasat - \gammabar$}, and $\sigma_\text{b}(\omega)$ represents a fixed background of Lorentz oscillators.

\begin{figure}[tbhp]
\begin{center}
\includegraphics[width=0.85\columnwidth]{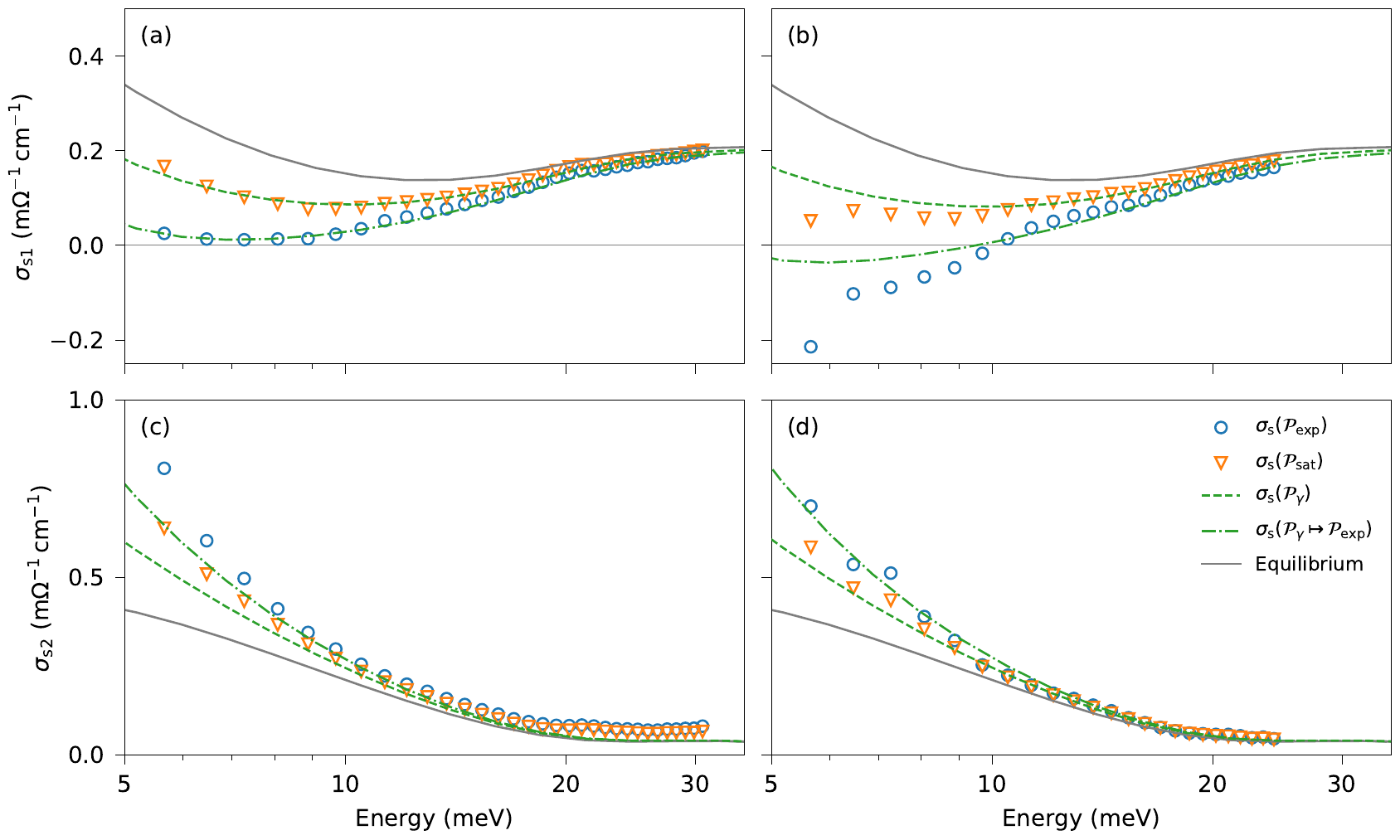}
\caption{Fit to measurements at \mbox{$F = 3.0~\funit$}~(a,c)~\cite{budden2021} and \mbox{$F = 4.5~\funit$}~(b,d)~\cite{buzzi2021} with \pgamma. Dashed and dot-dashed lines show  \mbox{$\sigmasurf(\omega, f; \pgamma)$} and \mbox{$\sigmasurf(\omega, f; \pgamma\mapsto\pexp)$}, respectively. Real (a,b) and imaginary (c,d) parts of \sigmasurf\ with different profile assumptions are reproduced from Fig.~4 and Fig.~5 of the primary text. Solid lines show $\sigmabar(\omega)$.}
\label{fig:joint-gamma}
\end{center}
\end{figure}

Using a procedure similar to the one we used to specify the parameters of \psat, we get \mbox{$\fgamma = 0.6~\funit$} and \mbox{$\gammasat = 0.85$~meV}. Figure~\ref{fig:joint-gamma} shows $\sigmasurf(\omega, f; \pgamma)$ and \mbox{$\Delta\sigmasurf(\omega, f; \pgamma\mapsto\pexp)$} at \mbox{$F = 3.0~\funit$} and \mbox{$F = 4.5~\funit$}. The associated surface relaxation rates  are \mbox{$\hbar\gammasurf = 1.3$~meV} at $F = 3.0~\funit$ and \mbox{$\hbar\gammasurf = 1.2$~meV} at $F = 4.5~\funit$. 
%
%

\section{Field and fluence dependence}\label{sec:field-fluence-dep}
\begin{figure}[tbhp]
\begin{center}
\includegraphics[width=0.7\columnwidth]{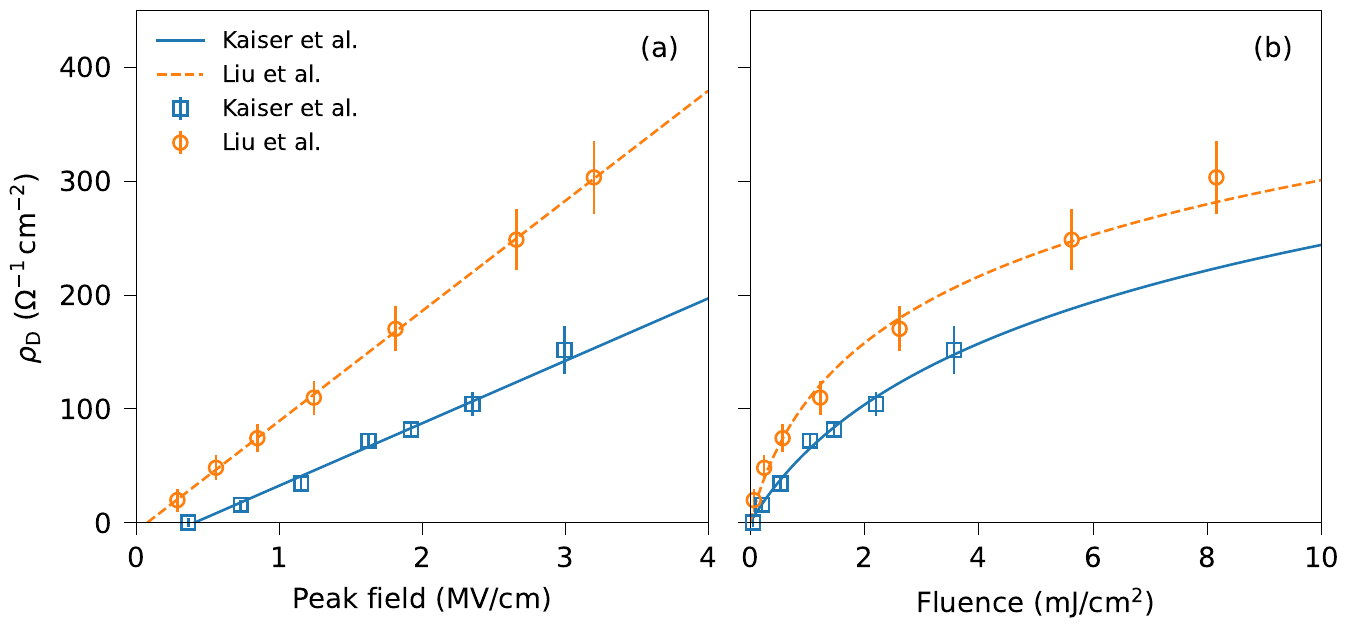}
\caption{Photoexcited Drude weights obtained from \mbox{\textcite{kaiser2014}} and \mbox{\textcite{liu2020}} as a function of field~(a) and fluence~(b) for \ybco\ at \mbox{$T = 100~\text{K}$} with pump frequency $\nu\approx 20~\text{THz}$. Lines show linear fits to the field dependence~(a) and fits with Eq.~\eqref{eq:fluence-dep} to the fluence dependence~(b). Uncertainty bounds for \mbox{\textcite{liu2020}} were not reported for these measurements, so we add a 10\% multiplicative uncertainty in quadrature with a $10~\Omega^{-1}\,\text{cm}^{-2}$ additive uncertainty, which is consistent with similar measurements from Ref.~\cite{liu2020}.} 
\label{fig:field-fluence}
\end{center}
\end{figure}

In measurements on \ybco, \mbox{\textcite{kaiser2014}} and \mbox{\textcite{liu2020}} reported an empirical linear relationship between the photoexcited Drude weight,
\begin{equation}
\rho_\text{D} = \epsilon_0\omegap^2 = \lim_{\omega\rightarrow 0}\omega\sigmasurfim(\omega; \pexp),
\label{eq:drude-weight}
\end{equation}
and the peak electric field $E_0$ of the pump, as shown in Fig.~\ref{fig:field-fluence}(a), and they interpreted $\rho_\text{D}$ as a measure of the superfluid stiffness of a photoinduced superconductor. The slope observed by \mbox{\textcite{liu2020}} is about a factor of two larger than that observed by \mbox{\textcite{kaiser2014}} on the same material and at the same temperature. But the pump pulse duration $\tau_\text{p}$ used by \mbox{\textcite{liu2020}} was a factor of two larger than that used by \mbox{\textcite{kaiser2014}}, so when the measurements are represented as a function of fluence $F = E_0^2 \tau_\text{p}/2Z_0$ they only differ by about 30\%, as shown in Fig.~\ref{fig:field-fluence}(b). The measurements are also consistent with
\begin{equation}
\rho_\text{D} = \rho_{\text{D}0}\ln(1 + F/\fsat),
\label{eq:fluence-dep}
\end{equation}
which follows from Eq.~(7) of the main text. Note that a strict proportionality between $\rho_\text{D}$ and $E_0$ would imply $\rho_\text{D}\propto\sqrt{F}$, so the proposed photoinduced superconducting state would be infinitely susceptible to the pump fluence. However, the intercepts are negative in both linear fits to $\rho_\text{D}(E_0)$, as one would expect if $\rho_\text{D}(F)$ were linear in the limit $F\rightarrow 0$.

\bibliography{pisc}